\documentclass{article}
\usepackage[utf8]{inputenc}
\usepackage{a4,latexsym,graphicx,color}
\usepackage{amsmath}
\usepackage{subcaption}

\usepackage[numbers,sort&compress]{natbib}

\newcommand{\dd}[1]{{\rm{d}}{#1}}

\title{Prediction of the capillary pressure of fluid surrounding a cylinder representing an idealized rock structure in porous media}

\author{Afshin Davarpanah\footnote{Correspondence to: afd6@aber.ac.uk}, Simon Cox, \\
Department of Mathematics, Aberystwyth University,\\ Aberystwyth SY23 3BZ, United Kingdom}

\date{}

\begin{document}

\maketitle

\begin{abstract}
Liquids in oil-bearing porous media assume complex shapes that depend on the reservoir characteristics and the wetting properties of the liquid. The wide variation in the geometry of rock formations makes it difficult to accurately predict the capillary pressure of small volumes of liquid and hence the likelihood of being able to move it. Here we consider the situation in which a small volume of liquid surrounds an upright cylinder on a flat substrate and predict the shape that the liquid takes and its capillary pressure. We validate our predictions by comparing with Surface Evolver simulations for a range of contact angles and cylinder radii. 
\end{abstract}


\section{Introduction}

The presence of small volumes of a liquid phase in a porous medium is common in processes such as  enhanced oil recovery~\cite{lake}, stimulation of plugged wells, environmental remediation~\cite{cantatbook}, and drying via evaporation~\cite{chen}. Capillary forces can cause the spontaneous imbibition of the liquid phase into micro or nano-sized pores in porous media, disrupting flow and affecting the efficiency of these processes. Arrays of cylindrical pillars are often used as a canonical example of a microfludic porous medium~\cite{chen,horner,byonkim}, with the aim of optimizing parameters such as the pressure drop required to mobilize the liquid. It is therefore necessary to have a good understanding of the distribution of liquid in the medium, i.e. both its location and its geometry. 
 
The most important parameter governing the feasibility of, say, oil recovery is the capillary pressure ($p_c$).  This is the difference in pressure between two phases coexisting in the porous medium and depends upon {\em inter alia}, each phase's volume, relative surface tensions (or contact angles), and pore geometry~\cite{hassanizadeh,soligno,wong13,rowlinson,xuxiao}. When regions of different fluids meet, the capillary pressures must equilibrate, and this will determine if the fluid will remain trapped in the complex geometry or whether it can be recovered.  

Wetting behavior is described by the contact angle at which interfaces meet, for example between the solid and the liquid and between the air and the liquid. Small contact angles correspond to wetting liquids, with a strong affinity for the substrate, while large contact angles are associated with non-wetting, or hydrophobic, surfaces~\cite{degennes}. Therefore, this parameter has a significant effect on the shape of liquid interfaces in porous media.

The internal geometry of porous media itself may also be difficult to examine and to characterize.  For this reason, many studies now turn to manufactured microfluidic geometries to test theories against experiments~\cite{horner,byonkim,xiao,jeong,cuib17,haghighi}. One element of such a porous medium might be a solid inclusion that spans the depth of the geometry. For simplicity, we consider such an idealized situation here, which allows is to make progress in developing a predictive model for the capillary pressure of a liquid in contact with such an inclusion. We derive an approximate solution of the Young-Laplace Law~\cite{soligno} for the capillary pressure of liquid surrounding a cylindrical inclusion, as studied experimentally by Chen et al.~\cite{chen} in the context of evaporation, and compare this with accurate Surface Evolver simulations to determine its range of validity. We explore in particular the effect of different contact angles on the shape and extent of this meniscus for a range of liquid volumes.

The techniques that we use in our derivation are those used to establish the shapes of bubbles~\cite{teixeira15}, 
menisci~\cite{howell99} and other liquid interfaces in 
soft matter~\cite{paunov92}. More usually, the interface rests on a liquid surface which also 
deforms~\cite{teixeira15,pozrikidis11}, and it appears that the case considered here, of a liquid meniscus on a solid substrate~\cite{teixeira18}, has received less attention.

 \section{Geometry and methods}
\subsection{Geometry}

\begin{figure}
\centering
\begin{subfigure}[b]{0.54\textwidth}
  \centering
  \includegraphics[width=\textwidth]{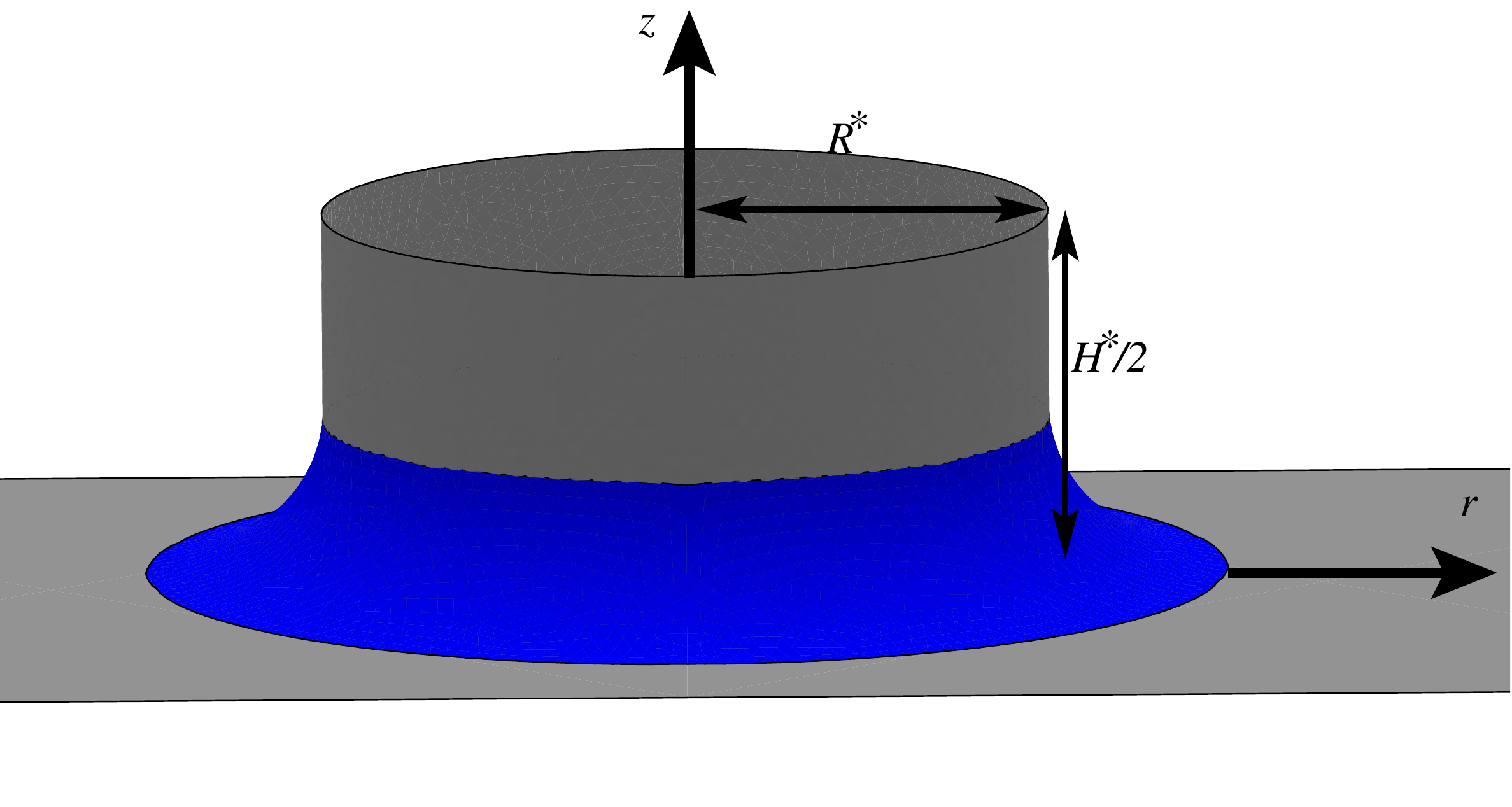}
    \caption{}
\end{subfigure}
\begin{subfigure}[b]{0.42\textwidth}
  \includegraphics[width=\textwidth]{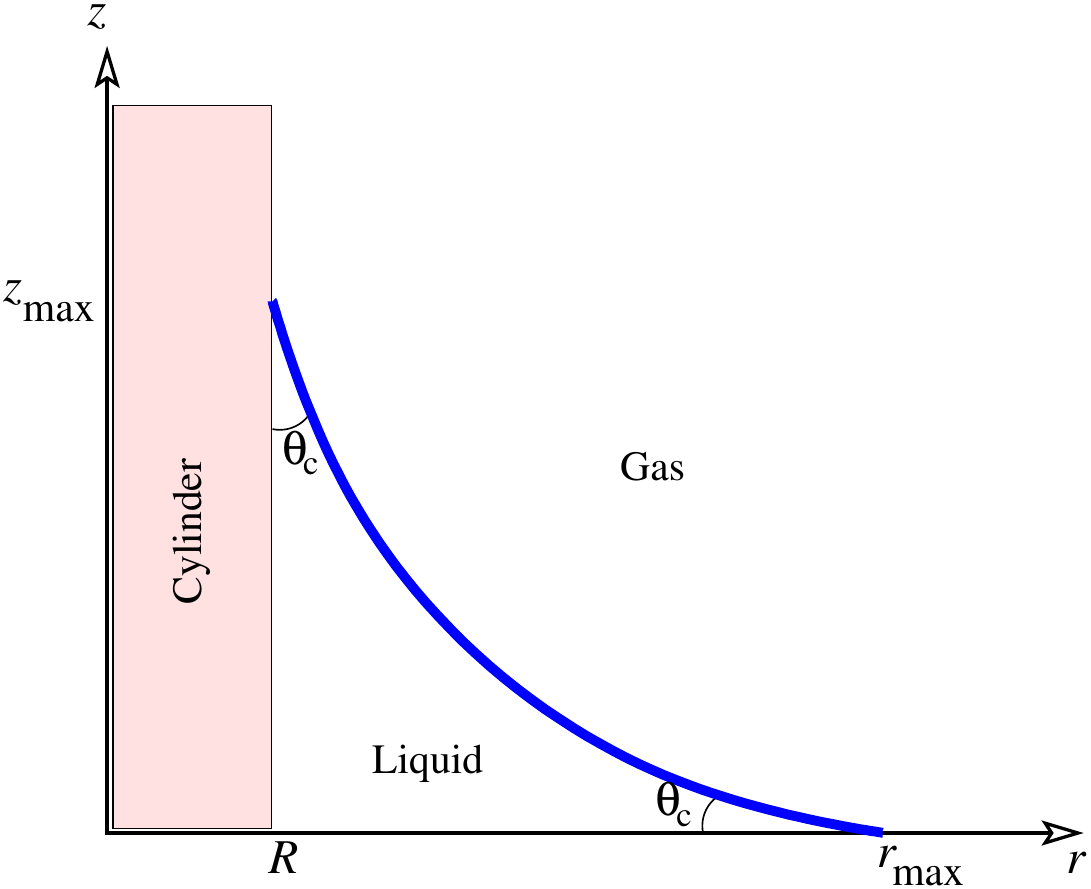}
    \caption{}
\end{subfigure}
\caption{We seek the shape of a liquid meniscus around the base of a vertical cylinder with circular cross-section. (a) 3D view, showing the lower half of the cylinder of radius $R^\ast$ and height $H^\ast$, representing an idealized rock structure in a porous medium, and the liquid meniscus in blue. The wetting films on the cylinder and on the substrate are not shown. (b) Since the solution is axisymmetric, it can be described by (dimensionless) radial position $r$ and vertical position $z$, and we consider only the $r-z$ plane. We consider the case where the contact angles $\theta_c$ on the cylinder and the substrate are equal. In this plane the interface has radius of curvature $r_c$, surface tension $\gamma$, extent $r_{\rm max}$ in the radial direction, and height $z_{\rm max}$.}
\label{fig:sketch}
\end{figure}

Our idealized porous medium consists of an upright circular cylinder with radius $R^\ast$ spanning the gap between two flat horizontal parallel surfaces with separation $H^\ast$. The gap between the surfaces is filled with air except for a narrow meniscus of liquid around the top and bottom of the cylinder. We assume that the gap between the surfaces is smaller than the capillary length, so that we can neglect the effects of gravity. Since we expect the size of rock pores to be of order microns while the capillary length is of order millimeters, this is likely to be a good approximation. Then by symmetry, we need only consider the lower half of this geometry, as shown in Figure~\ref{fig:sketch}, and at such a scale, we expect capillary pressures to be of order $\gamma/H^\ast$, where $\gamma$ is the interfacial tension, of order $30 \times 10^{-3}$ N/m. 

As the liquid volume varies, the capillary pressure will vary from small values at large volumes, where the radius of curvature in the $(r,z)$ plane is large, to much larger values at small volumes. Where the liquid meets the cylinder and where it meets the plate it does so at a particular angle, the contact angle. The contact angle is affected by the properties of the liquid and the surfaces it contacts, so it could be different on the cylinder and the substrate. For simplicity, we take these two values to be the same, but the results described below are straightforward to generalise to the case where they are different.

 The contact angle is directly related to the balance between the surface tension  of the liquid-air interface and the surface tension  of the wetting film on the solid walls. If we denote the contact angle by $\theta_c$ then                           
 \begin{equation}
    \cos \theta_c = \frac{\gamma}{\gamma_w},
\label{eq:contactangle}
\end{equation} 
as shown in Figure ~\ref{fig:sketch}(b).

The shape of the interface is given by a solution of the Young-Laplace Law~\cite{soligno,yeung97}, which balances the capillary pressure with the local mean curvature at each point of the interface. In this cylindrical geometry, the expression for the curvature is highly non-linear: it has two components, consisting of the principal radii of curvature in two perpendicular directions, and these both vary with vertical position within the meniscus. Our prediction of the capillary pressure will therefore be an approximation, which performs best for large rock structures within the porous medium ($R^\ast \gg H^\ast$), and so to determine how accurate our prediction is we will compare it with precise numerical solutions. 

\subsection{Numerical method}

We work in dimensionless coordinates, with all lengths scaled by the gap height $H^\ast$. We define the dimensionless cylinder radius to be $R =R^\ast /H^\ast$ and choose (without loss of generality) a dimensionless surface tension $\gamma = 1$. Then a dimensionless capillary pressure of $p_c = 1$ in our results below corresponds to a dimensional capillary pressure of $\gamma/H^\ast = 30 \times 10^{-3} / 10^{-6} = 10^3 {\rm Pa}$.

To perform the simulations we use Surface Evolver~\cite{brakke92}, which finds the shape of fluid interface shapes with given surface tension and other energies under various constraints. Similar calculations have been performed for small pockets of liquid trapped between spheres~\cite{hilden03}. The situation that we consider is axisymmetric so that we need only consider the shape of the interface in the $(r,z)$ plane (see Figure~\ref{fig:sketch}), which consists of a curve connecting the cylinder to the substrate with the given contact angle at each end and a specified enclosed volume.

This curve is discretized into roughly $2^5$ straight elements. The contact angles are considered as further energies, and the system is driven to a minimum of energy using a gradient descent method. In the simulations it is straightforward to vary the cylinder radius $R$, the liquid volume $V$ and the contact angle $\theta_c$ to explore all relevant parameters.

\subsection{Mathematical Models}

In this relatively simple geometry, we can derive an approximate formulae for the capillary pressure. The Young-Laplace equation for an interface at equilibrium without gravity is
\begin{equation}
    p_c = p_{air} - p_{liquid} = \gamma C,
\end{equation}
in which the mean curvature $C$ of the interface can be written~\cite{teixeira15}: 
\begin{equation}
C = -\left[ 1+ r_z^2\right]^{-\frac{3}{2}}\left( r_{zz} - \frac{1+r_z^2}{r} \right) ,
\label{eq:curvature}
\end{equation}
where subscripts denote derivatives. There are two terms here, corresponding to the two radii of curvature of the interface, one around the cylinder and the other perpendicular to it in the $r-z$ plane.
Writing $\hat{p}_c$ for $p_c/\gamma$ leads to the following equation for the capillary pressure:
\begin{equation}
\hat{p}_c = \left[ 1+ r_z^2\right]^{-\frac{3}{2}}\left( r_{zz} - \frac{1+r_z^2}{r} \right).
\label{eq:LY1}
\end{equation}

Equation (\ref{eq:LY1}) is too complex to find a closed-form solution. Instead, we assume that in typical porous media the size of the rock structures is large, $R \gg 1$. We will consider two approximate solutions: the first will neglect the second term entirely, while the second will approximate it.

\subsubsection{Approximation 1: interface shape in the limit of large cylinder radius}

The second term in the parentheses in eq.(~\ref{eq:LY1}), containing the factor $1/r$, will be dominated by the first at large $R$, and in our first approximation we choose to neglect it. As we shall see below, we are effectively approximating the interface as a circular arc (precisely one-quarter of a circle if $\theta_c = 0$) in the $(r,z)$ plane.

Equation (\ref{eq:LY1}) becomes
\begin{equation}
    \hat{p}_c = {\rm sign}\left(\frac{\pi}{4} - \theta_c \right)
    \frac{r_{zz}}{\left[ 1+ r_z^2\right]^{\frac{3}{2}}},
    \label{eq:LYapprox1}
\end{equation}
in which ${\rm sign}\left(\theta_c - \pi/4\right)$ reflects the fact that, in this approximation, the curvature of the interface will change sign at a contact angle of $\pi/4$. That is, the capillary pressure will go from positive to negative, i.e., the liquid pressure will exceed the air pressure. 

Following Teixiera et al.~\cite{teixeira15}, we introduce a coordinate $\theta$ with $\cot\theta = r_z$  in Eq.~(\ref{eq:LYapprox1}) to give 
\begin{equation}
  \hat{p}_c = {\rm sign}\left(\frac{\pi}{4} - \theta_c \right) \sin\theta\frac{\dd{\theta}}{\dd{z}}.
\label{eq:LYthetaapprox1}
\end{equation}
Integrating gives the height of the interface in terms of the parameter $\theta$:
\begin{equation}
    z(\theta) = \hat{p}_c \left( \cos \theta_c -\cos\theta \right)
    \quad \mbox{with} \quad \theta \in \left( \theta_c, \frac{\pi}{2} - \theta_c \right),
    \label{eq:ztheta1}
\end{equation} 
although $\hat{p}_c$ remains unknown at this stage. The maximum height reached by the meniscus on the cylinder occurs when  $\theta = \pi/2-\theta_c$:
\begin{equation}
    z_{\rm max} = \left( \cos \theta_c - \sin\theta_c \right) / \hat{p}_c.
    \label{eq:zmax}
\end{equation}
This expression is shown in Figure 5 below, which we describe once the value of $\hat{p}_c$ is predicted.

Since the interface is now approximated by an arc of a circle, the maximum extent of the meniscus in the radial direction is 
\begin{equation}
    r_{\rm max} = R + \left( \cos \theta_c - \sin\theta_c \right)/ \hat{p}_c .
    \label{eq:rmax}
\end{equation}

We now replace $\theta$ from Eq.~\ref{eq:LYthetaapprox1} using $\cot\theta = \displaystyle \frac{\cos\theta}{\sqrt{1-\cos^2\theta}}=-r_z$ to find the interface profile $z(r)$. 
We substitute for $\cos\theta$ from eq.~(\ref{eq:ztheta1}) and integrate
\begin{equation}
    -\frac{\dd{r}}{\dd{z}} = \frac{\cos\theta_c -\hat{p}_c z}{\sqrt{1-(\cos\theta_c - \hat{p}_c z)^2}}
\end{equation}
to obtain
\begin{equation}
    \left( z - \frac{\cos\theta_c}{\hat{p}_c} \right)^2 + 
    \left( r - \left(R+\frac{\cos\theta_c}{\hat{p}_c}\right)  \right)^2
     = \frac{1}{\hat{p}_c^2}
     \label{eq:circle_approx1}
\end{equation}
Clearly, this is the equation of a circle with radius $1/\hat{p}_c$  and centre $(R+r_c,r_c)$, where  $r_c =  \cos\theta_c/\hat{p}_c$. The radius of curvature increases, and the meniscus gets larger, as the capillary pressure decreases, as expected. As the contact angle changes the centre of curvature moves closer to the axes. We have yet to determine the effect on the capillary pressure of changes in the contact angle, which we do next.

\subsubsection{Approximation 1: capillary pressure in the limit of large cylinder radius}

Let us assume that $R$, $\theta_c$ and $\gamma$ are given. We need to find the capillary pressure $\hat{p}_c$  to completely solve the problem, which we do by determining the liquid volume
\begin{equation}
    V = \int_0^{r_c(1+\tan\theta_c)} z(r) r \dd{r}
\end{equation}
with z(r) from Eq.~(\ref{eq:circle_approx1}). 
The substitution $r= R+ r_c \left(1-\frac{\cos\theta}{\cos\theta_c}\right)$  allows us to resolve the integral:
\begin{equation}
    V = 2 \pi r_c^2 \left\{ R 
    \left[ 1-\tan\theta_c-\frac{\frac{\pi}{4}-\theta_c}{\cos^2\theta_c}\right]
    + r_c 
    \left[ \frac{(1-\tan\theta_c)^2}{2} -\frac{\frac{\pi}{4}-\theta_c}{\cos^2\theta_c} + \frac{1-\tan^3\theta_c}{3} \right] \right\}.
\end{equation}
In keeping with our assumption that $R$ is large, we retain only the first term, proportional to $r_c^2 R$, and neglect the second term proportional to $r_c^3$:
\begin{equation}
    V = 2 \pi r_c^2 R 
    \left[ 1-\tan\theta_c-\frac{\frac{\pi}{4}-\theta_c}{\cos^2\theta_c}\right]
    =  2 \pi \hat{p}_c^2 R 
    \left[ \cos^2\theta_c -\frac{1}{2}\sin(2\theta_c)+\theta_c-\frac{\pi}{4} \right].
\end{equation}
Introducing the function $f(\theta_c) = \cos^2\theta_c -\frac{1}{2}\sin(2\theta_c)+\theta_c-\frac{\pi}{4}$ for the dependence on the contact angle then gives
 \begin{equation}
     \hat{p}_c \approx {\rm sign}\left( \frac{\pi}{4} - \theta_c\right)
        \sqrt{ \frac{2\pi R f(\theta_c)}{V} }.
    \label{eq:volapprox1}
 \end{equation}
This dependence on contact angle precisely matches the expression for a straight channel given by Ma et al.~\cite{mamm96}, who relate the capillary pressure to the area of the cross-section of the channel occupied by liquid. The advantage of our expression is that it allows us to relate the capillary pressure to the volume of the cylindrical meniscus, $\hat{p}_c(V)$, rather than its cross-sectional area (which is more difficult to determine in an experiment).

\subsubsection{Approximation 2: Interface shape dependent on cylinder radius}

In the preceding approximation we completely neglected the second radius of curvature in the Laplace Young law to find a closed-form solution of eq.~(\ref{eq:LY1}) in the shape of an inverted circle. We now improve upon this prediction by introducing a dependence on the cylinder radius $R$. We again assume that $R$ is large and replace the radial position $r$  in the denominator of the second term in eq.~(\ref{eq:LY1}) with the cylinder radius $R$ to give:
\begin{equation}
\hat{p}_c = \left[ 1+ r_z^2\right]^{-\frac{3}{2}}
    \left( r_{zz} - \frac{1+r_z^2}{R} \right).
\label{eq:LY2}
\end{equation}
This leads to a semi-analytic expression for the quantities of interest. (It is semi-analytic because we have to solve for the capillary pressure numerically, even though we will find closed-form expressions for the shape and extent of the meniscus.)

Proceeding as before, and assuming for the moment that $\hat{p}_c$ is positive, we introduce the coordinate $\theta$: eq.~(\ref{eq:LY2}) becomes
\begin{equation}
\hat{p}_c = \sin\theta \left( \frac{\dd{\theta}}{\dd{z}} - \frac{1}{R} \right),
\label{eq:LYtheta}
\end{equation}
which improves upon eq.~(\ref{eq:LYthetaapprox1}).
We rearrange this into the more convenient form 
\begin{equation}
\frac{\dd{z}}{\dd{\theta}} = \frac{R \sin\theta }{R \hat{p}_c+\sin\theta}.
\label{eq:LYthetaz}
\end{equation}
In the same way we have
\begin{equation}
\frac{\dd{r}}{\dd{\theta}} = -\frac{R \cos\theta }{R \hat{p}_c+\sin\theta}.
\label{eq:LYthetar}
\end{equation}
These two equations allow us to find the shape of the meniscus implicitly.

Of the two, eq.~(\ref{eq:LYthetar}) can be integrated more easily: 
using~\cite[\S 2.552.2]{gradsr} and evaluating the constant of integration at $r=R$ gives
\begin{equation}
r(\theta; R, \hat{p}_c, \theta_c) = R \left[ 1 + \ln \left(R \hat{p}_c + \cos \theta_c \right) - \ln \left(R \hat{p}_c + \sin\theta\right) \right].
\label{eq:rtheta}
\end{equation}
Hence the  maximum extent of the meniscus beyond $r=R$ is
\begin{equation}
r_{\rm max}(R, p_c, \theta_c) = R \left[ 1 + \ln \left(R \hat{p}_c + \cos \theta_c \right) - \ln \left(R \hat{p}_c + \sin\theta_c \right) \right],
\label{eq:rmaxtheta}
\end{equation}
which should be compared with eq.~(\ref{eq:rmax}).

We now integrate eq.~(\ref{eq:LYthetaz}) using~\cite[\S 2.551.2, 2.551.3]{gradsr}  and evaluate the constant of integration at $z=0$. There are now two branches of the solution depending on the value of $R\hat{p}_c$:
\begin{multline}
z(\theta; R, \hat{p}_c, \theta_c)  =  
     R \left[\theta-\theta_c\right] + \\
\left\{ 
  \begin{array}{ll}
     \frac{2 R^2 \hat{p}_c}{\sqrt{(R \hat{p}_c)^2-1}}
         \left[ \tan^{-1} \left( \frac{R\hat{p}_c \tan \left(\theta_c/2\right)+1} {\sqrt{(R \hat{p}_c)^2-1}}\right) -
           \tan^{-1} \left( \frac{R\hat{p}_c \tan \left(\theta/2\right)+1} {\sqrt{(R \hat{p}_c)^2-1}}\right)   \right] 
 & R \hat{p}_c > 1\\[2ex]
     \frac{2 R^2 \hat{p}_c}{\sqrt{1-(R \hat{p}_c)^2}}
         \left[ \coth^{-1} \left( \frac{R\hat{p}_c \tan \left(\theta/2\right)+1} {\sqrt{1-(R \hat{p}_c)^2}}\right) -
           \coth^{-1} \left( \frac{R\hat{p}_c \tan \left(\theta_c/2\right)+1} {\sqrt{1-(R \hat{p}_c)^2}}\right)   \right] 
 & R \hat{p}_c < 1.
   \end{array}
\right.
\label{eq:ztheta}
\end{multline}
This generalizes eq.~(\ref{eq:ztheta1}) to smaller $R$.
The height, $z_{\rm max}$, of the meniscus is then found by evaluating this expression with $\theta = \frac{\pi}{2}-\theta_c$.

In our first approximation the curvature of the interface changes sign when $\theta_c = \pi/4$. In this second approximation it is more difficult to predict when this happens since the magnitude of the first term in parentheses in eq.~(\ref{eq:LYtheta}) depends on not only the contact angle but also the liquid volume (and cylinder radius). However, only small changes to the expressions for $r(\theta)$ and $z(\theta)$, are required for them to continue to work: we change the signs in front of the two trigonometric functions in eq.~(\ref{eq:rtheta}), replace $+1$ with $-1$ in eq.~(\ref{eq:ztheta}), and take the absolute value of the capillary pressure in both formulae. In this way we can extend the prediction to negative $\hat{p}_c$, as shown in subsequent figures.

\subsubsection{Approximation 2: capillary pressure dependent on cylinder radius}

Before we can compare the interface shapes with the simulations, we need to determine the dependence of the capillary pressure on the volume $V$ of the meniscus (as well as on the parameters $R$ and $\theta_c$). This requires a further integration, for which we have not been able to determine a closed form expression. Instead, we choose to evaluate the following integral numerically:
\begin{equation}
V = 2 \pi \int_{\theta_c}^{\pi/2-\theta_c} z(\theta) r(\theta) \frac{\dd{r}}{\dd{\theta}} \dd{\theta},
\label{eq:vol1}
\end{equation}
substituting the appropriate expressions from eqs.~(\ref{eq:ztheta}), (\ref{eq:rtheta}) and (\ref{eq:LYthetar}) respectively. 

\begin{figure}
\centering
\begin{subfigure}{0.48\textwidth}
  \centering
  \includegraphics[width=\textwidth]{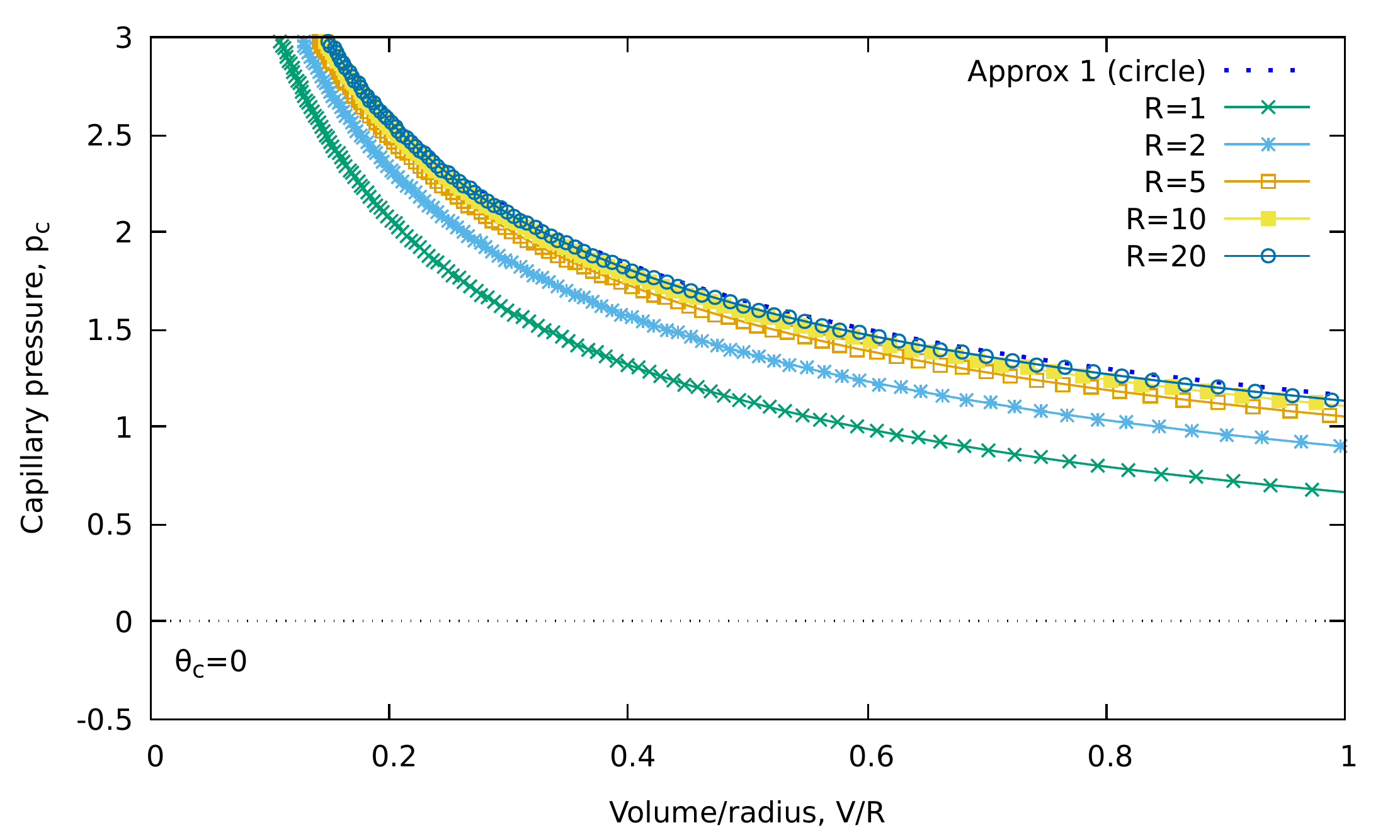}
  \caption{}
\end{subfigure}
\begin{subfigure}{0.48\textwidth}
  \includegraphics[width=\textwidth]{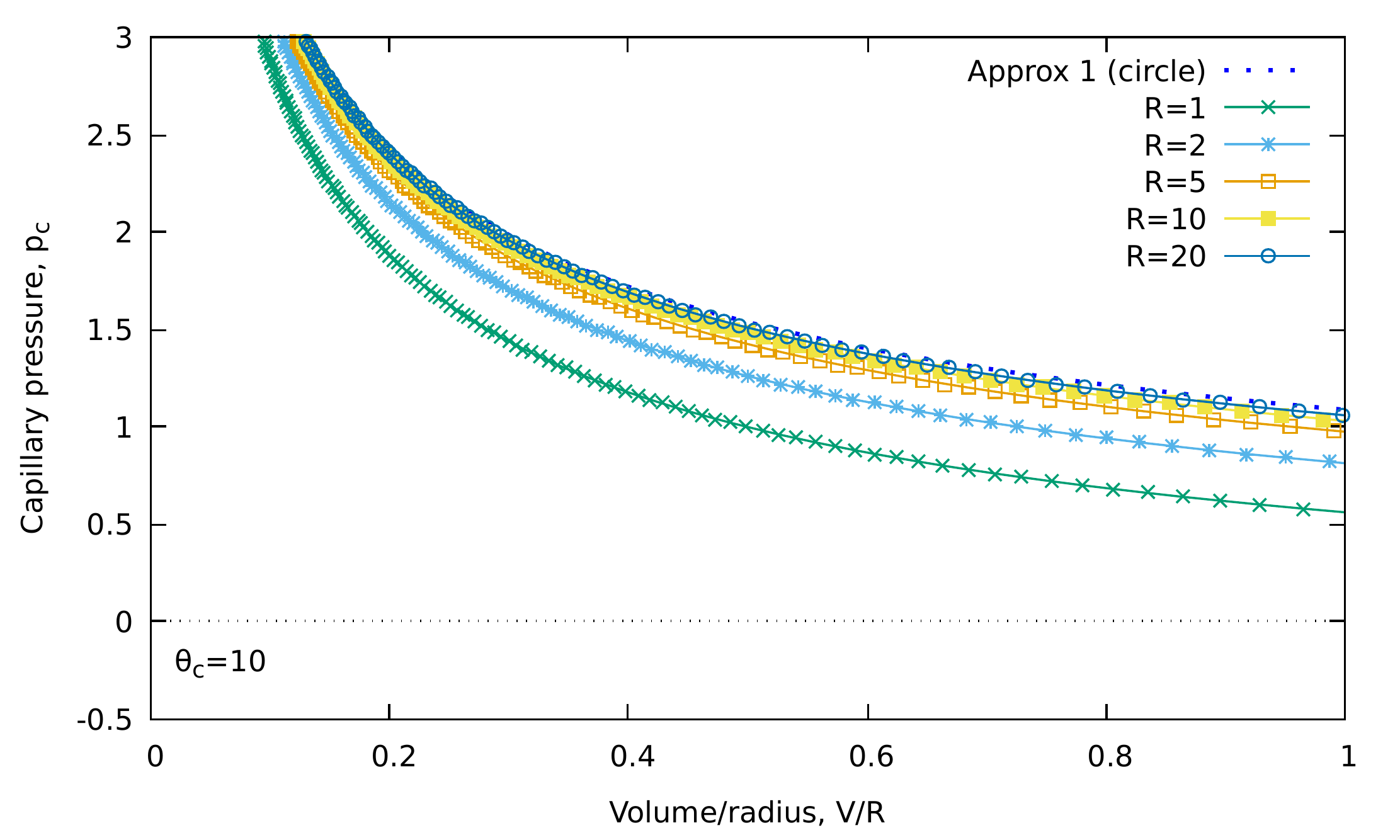}
    \caption{}
\end{subfigure}

\begin{subfigure}{0.48\textwidth}
  \centering
  \includegraphics[width=\textwidth]{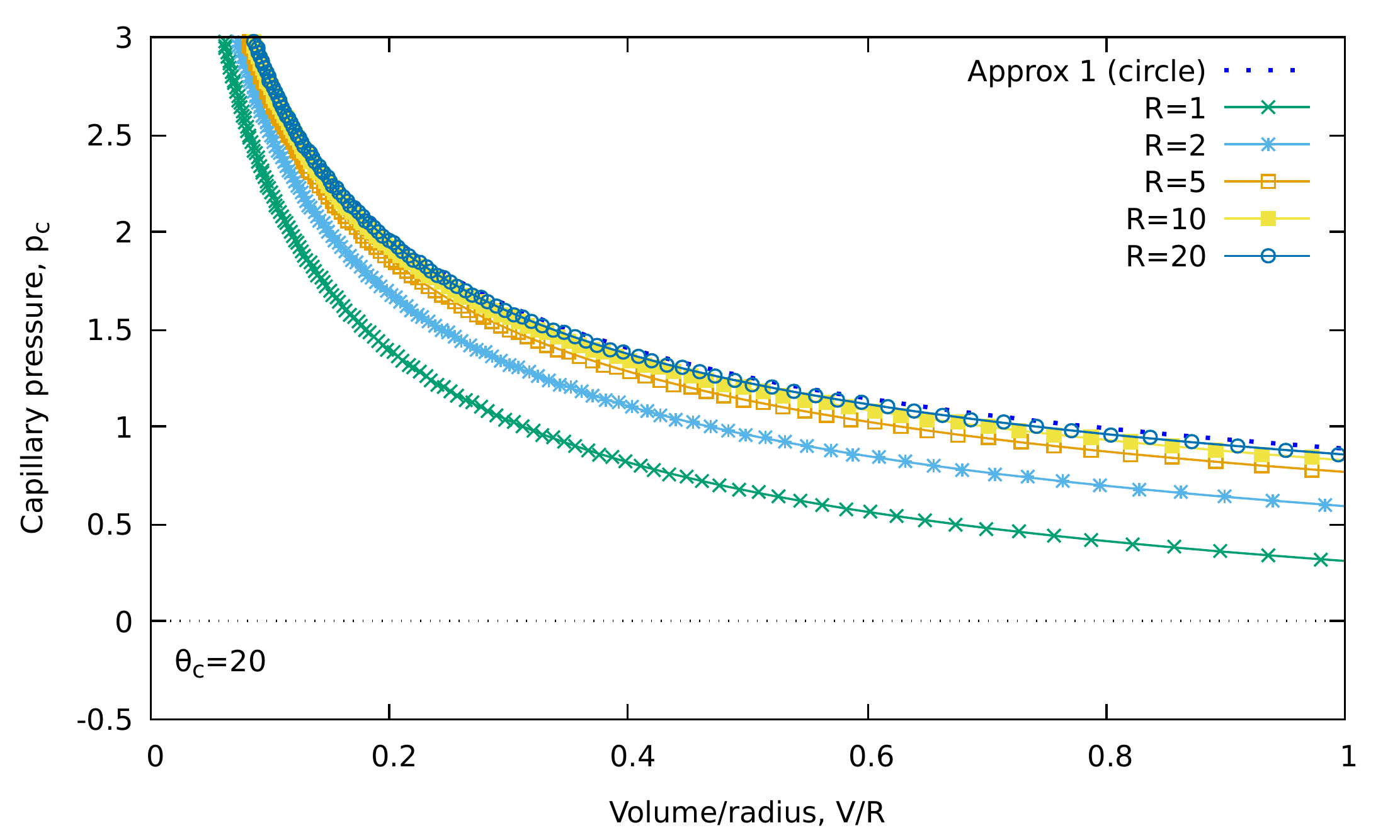}
    \caption{}
\end{subfigure}
\begin{subfigure}{0.48\textwidth}
  \includegraphics[width=\textwidth]{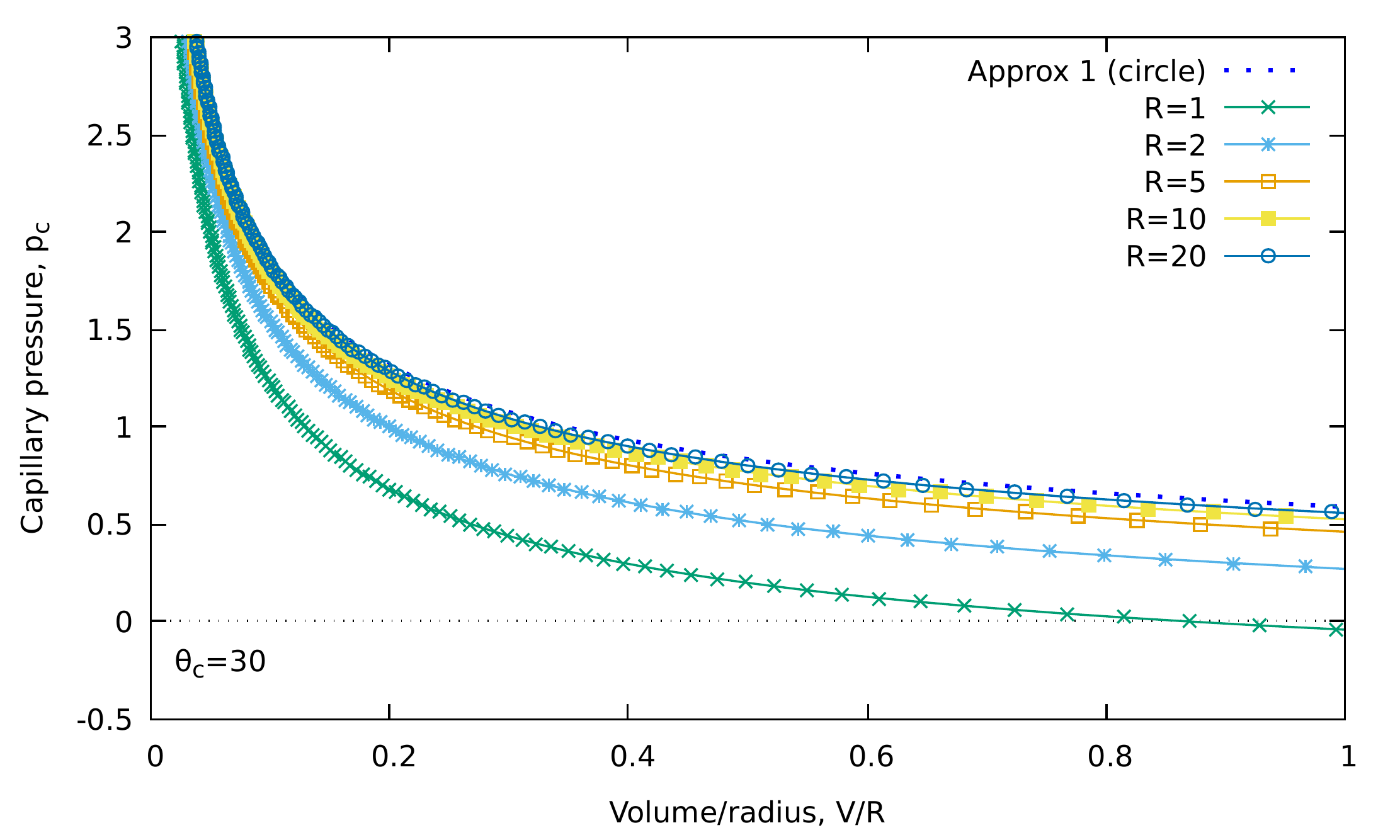}
    \caption{}
\end{subfigure}

\begin{subfigure}{0.48\textwidth}
  \includegraphics[width=\textwidth]{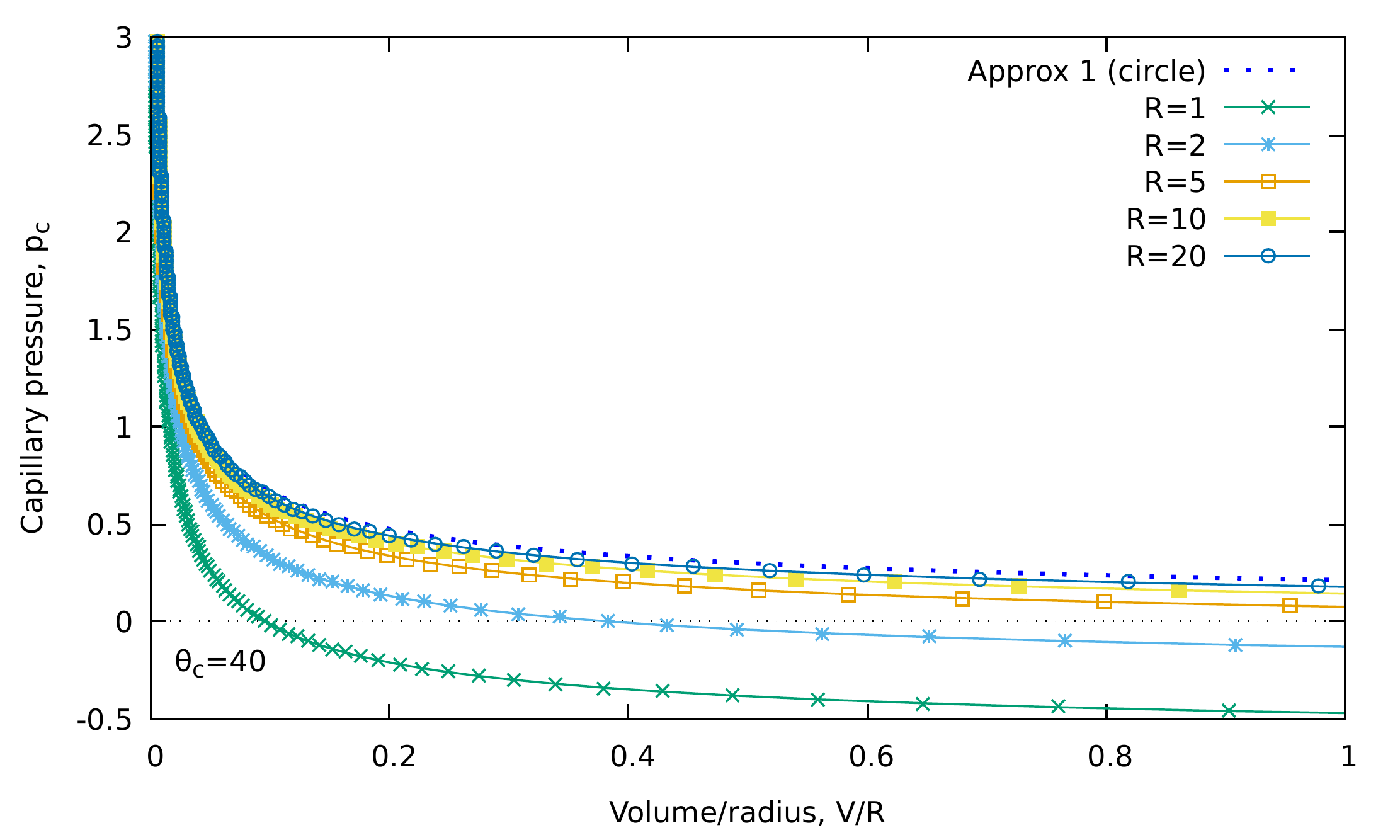}
    \caption{}
\end{subfigure}

\caption{Profiles of $\hat{p}_c$ against liquid volume for different values of the cylinder radius $R$. Each panel is for a different contact angle $\theta_c$. These are numerical integrations of eq.~(\ref{eq:vol1}), while the dotted line on each plot is the first approximation, eq.~(\ref{eq:volapprox1}). The horizontal axes are scaled by the cylinder radius $R$ to make the curves comparable.}
\label{fig:pc_vs_vol}
\end{figure}

The results of evaluating this integral allow us to determine a value for $\hat{p}_c$ corresponding to a particular liquid volume (for any given values of $R$ and $\theta_c$). The dependence of $\hat{p}_c$ on $V$ is shown in Figure~\ref{fig:pc_vs_vol}: a strong decrease from a large value for small $V$. We can now determine the interface shape, the extent of the meniscus in each direction ($r_{\rm max}$ and $z_{\rm max}$) and the dependence of each of these quantities on cylinder radius and contact angle. 

\section{Results}

 \subsection{Interface Shape}
 
 The volume of the meniscus is approximately its cross-sectional area multiplied by the circumference of the cylinder. Therefore, in the following, we consider liquid volumes that scale with $R$, with the expectation that the cross-section of the meniscus remains roughly constant in shape. We consider volumes $V$ that are small enough that $z_{\rm max} < H/2$ so that the liquid regions around the top and bottom of the cylinder never meet.
 
\begin{figure}
\centering
\begin{subfigure}[b]{0.48\textwidth}
  \centering
  \includegraphics[width=\textwidth]{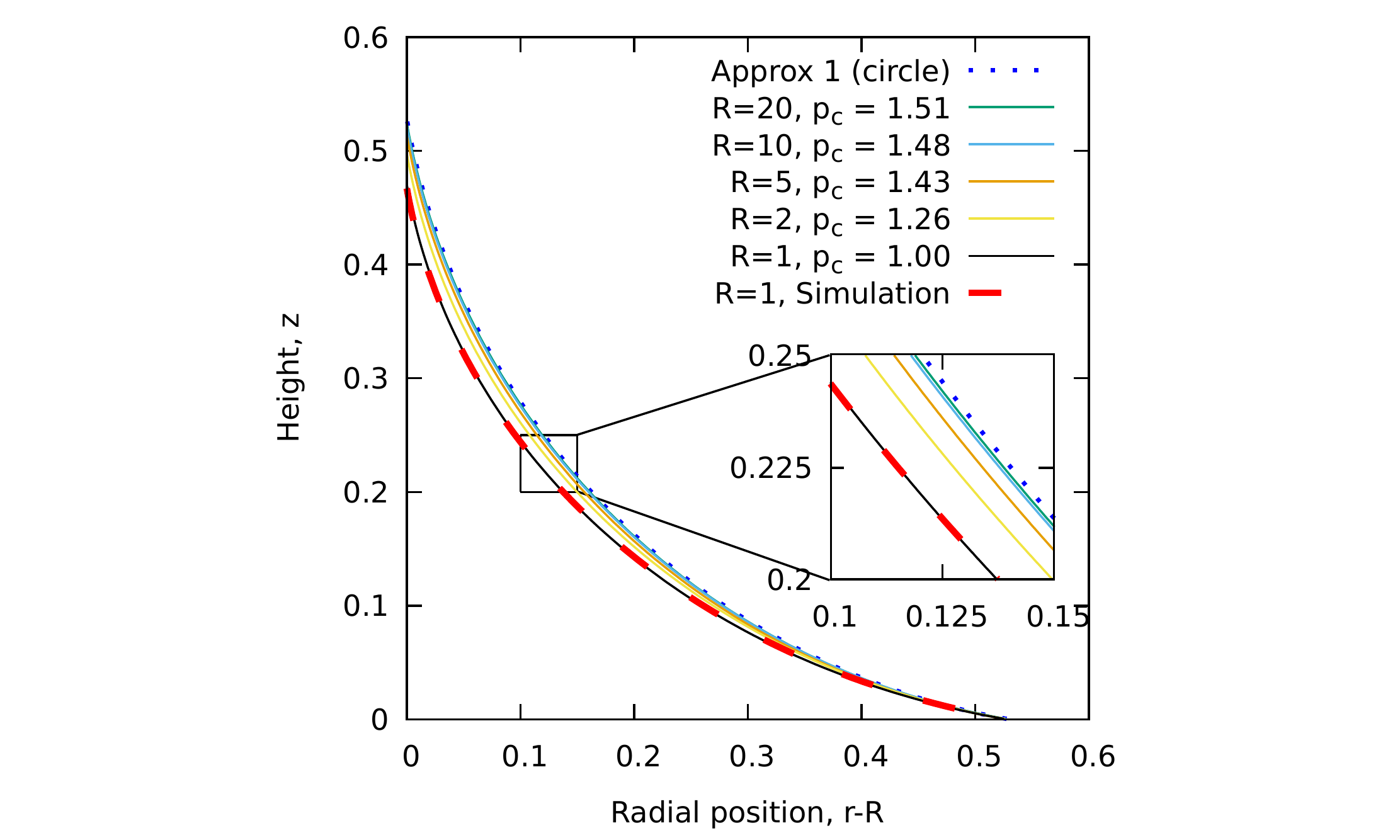}
    \caption{}
\end{subfigure}
\begin{subfigure}[b]{0.48\textwidth}
  \centering
  \includegraphics[width=\textwidth]{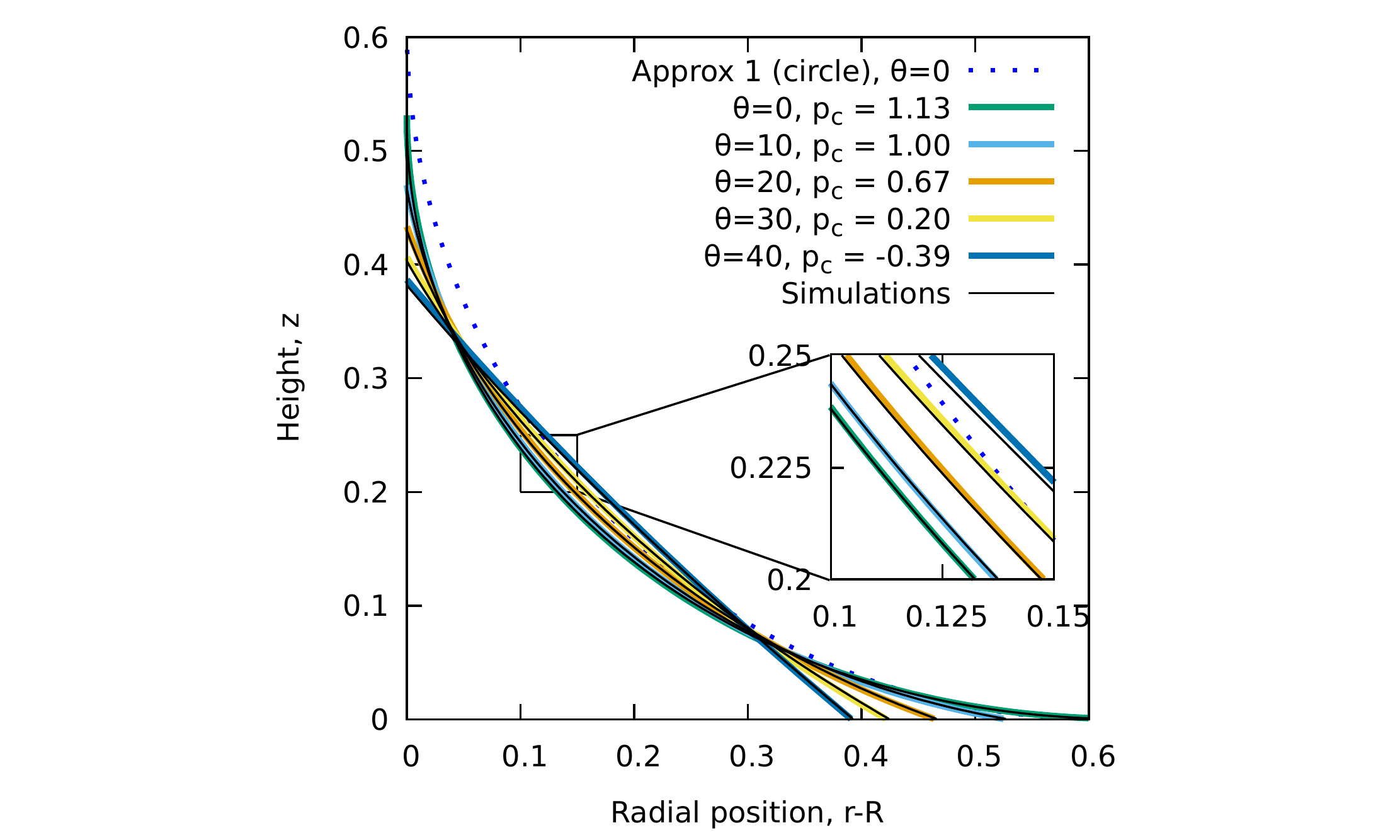}
    \caption{}
\end{subfigure}
\caption{Predicted interface shapes,  comparing a circle (our first approximation, eq.~(\ref{eq:circle_approx1}), dotted line), the second approximation (eqs.~(\ref{eq:rtheta}) and (\ref{eq:ztheta}), solid lines) and Surface Evolver simulations. In each case the liquid volume is $V=0.5R$ and the appropriate value of $\hat{p}_c$ for the second approximation is chosen from the profiles in Figure~\ref{fig:pc_vs_vol}. 
(a) Profiles in the $r-z$ plane for different cylinder radii $R$ and contact angle $\theta_c = \pi/18$. The simulation data for $R=1$ is shown as a dashed line. 
(b) Profiles in the $r-z$ plane for different contact angles $\theta_c$ and radius $R=1$. The simulation data is shown as solid lines that overlay the predictions. 
}
\label{fig:profile2}
\end{figure}

 The meniscus profile is found by plotting eq.(\ref{eq:ztheta}) against eq.(\ref{eq:rtheta}); an example is shown in Figure~\ref{fig:profile2}(a) in the case $\theta_c = \pi/18$ and $V=0.5R$. The interface does indeed become closer to a circle as $R$ increases, and the approximation eq.~(\ref{eq:circle_approx1}) for $R \rightarrow \infty$ works well for $R$ above about 5. The comparison of the second approximation with the Surface Evolver simulation for $R=1$ shows even better agreement, even though the premise for that approximation is that $R$ is large.

Similarly, we can determine the extent of the meniscus, i.e. values of $r_{\rm max}$ and $z_{\rm max}$ for different contact angles and cylinder radii. Examples are shown in Figure~\ref{fig:profile2}(b) and (c). 
As $R$ increases, the height of the meniscus increases towards the prediction of eq.~(\ref{eq:zmax}) until the circular shape is reached at large $R$, working well for $R$ greater than about 20.
More remarkable is that the radial extent of the meniscus $r_{\rm max}$ appears constant as the radius of the cylinder varies, much closer to the prediction of eq.~(\ref{eq:rmax}). However, the second approximation captures the variation of both $z_{\rm max}$ and $r_{\rm max}$ with $R$, except for a discrepancy in $r_{\rm max}$ at small $R$ for small contact angles.

\begin{figure}
\centering
\begin{subfigure}{0.48\textwidth}
  \centering
  \includegraphics[width=\textwidth]{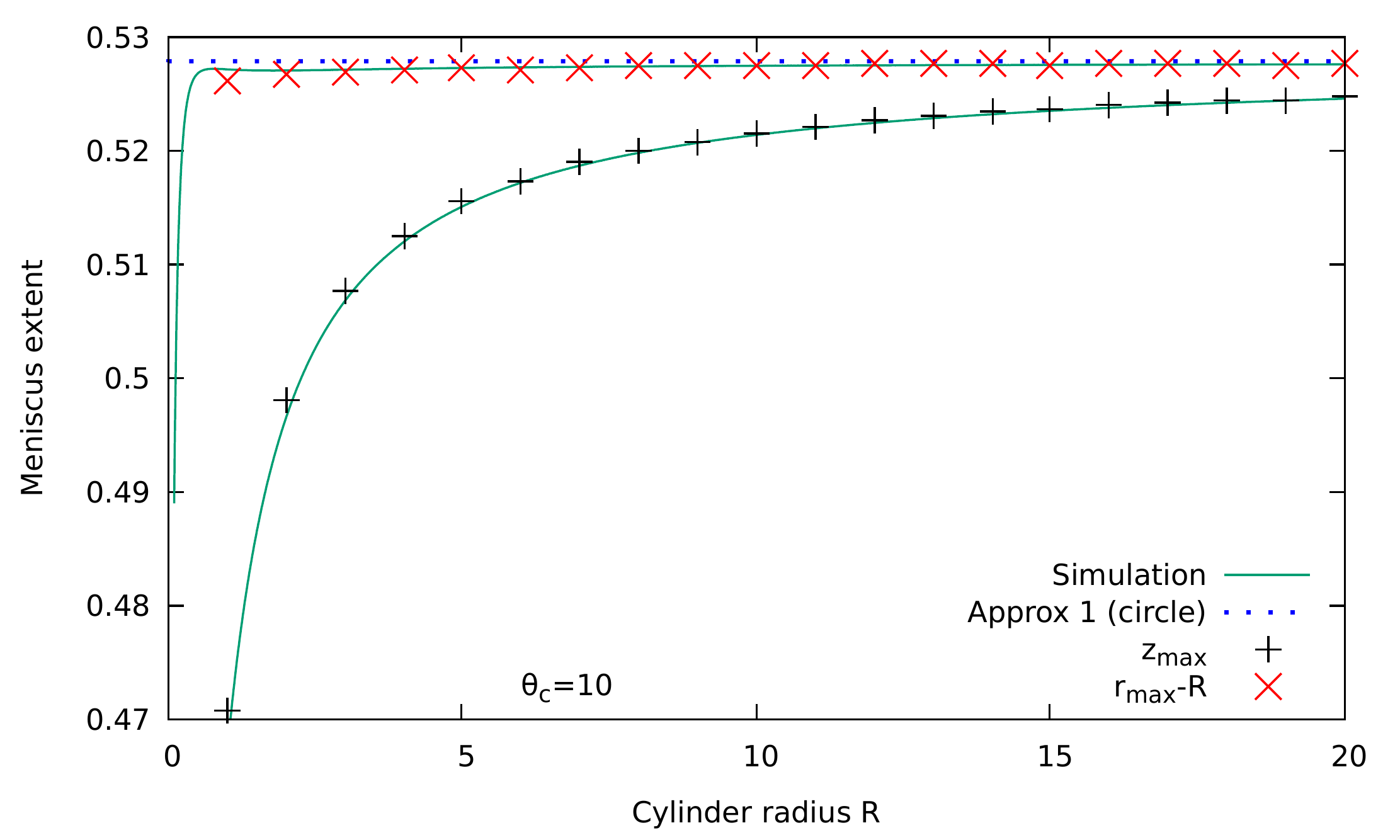}
    \caption{}
\end{subfigure}
\begin{subfigure}{0.48\textwidth}
\includegraphics[width=\textwidth]{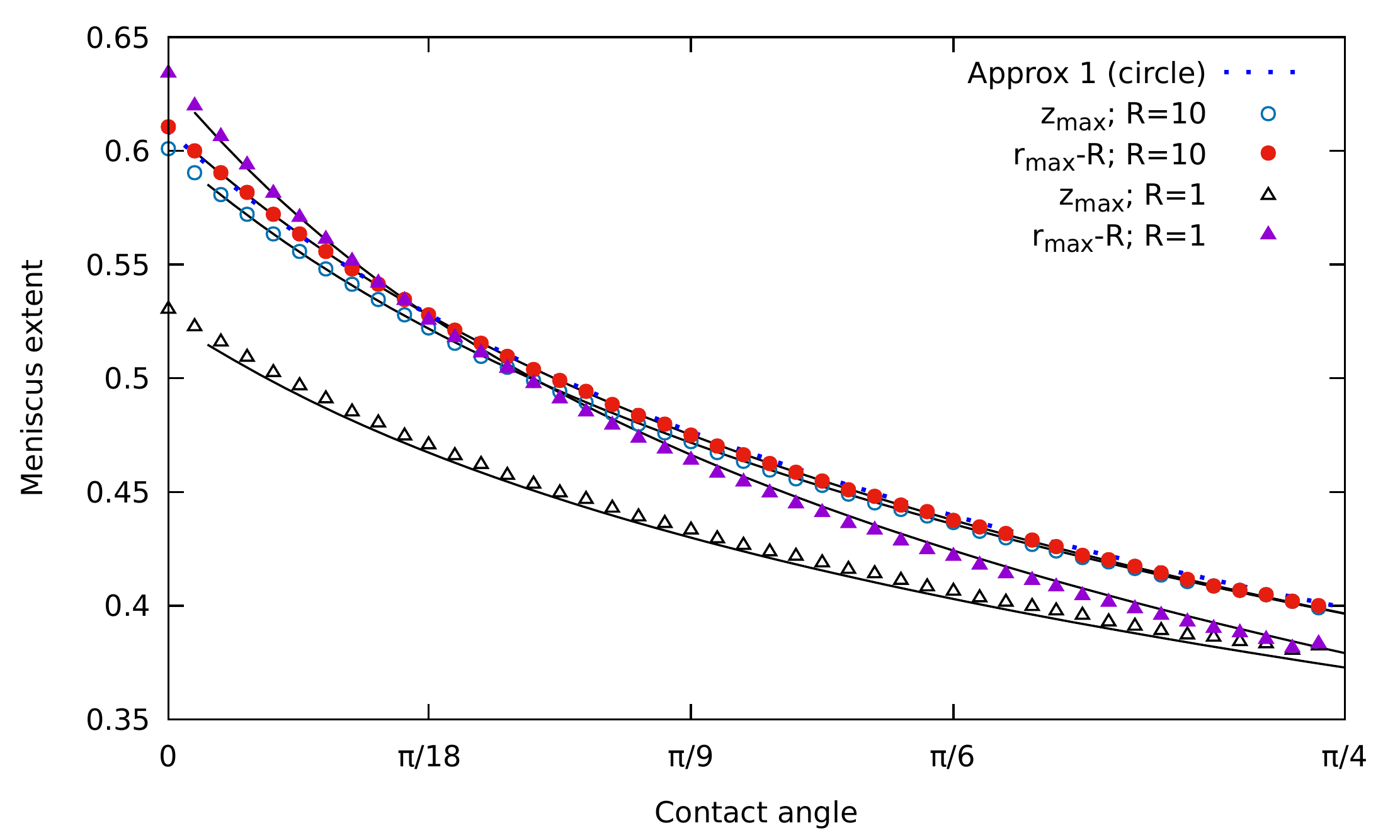}
    \caption{}
\end{subfigure}
\caption{Predicted values of the extent of the meniscus in the radial ($r_{\rm max}-R$) and vertical ($z_{\rm max}$) directions compared with Surface Evolver simulations. (a) With contact angle $\theta_c = \pi/18$ and liquid volume $V=0.5R$, as in Figure~\ref{fig:profile2}, the first approximation (eqs.~(\ref{eq:rmax}) and (\ref{eq:zmax}) with $\hat{p}_c$ from eq.~(\ref{eq:volapprox1}), shown as a horizontal dotted line) is independent of radius $R$, while the second approximation (eqs.(\ref{eq:rmaxtheta}) and (\ref{eq:ztheta}), shown as points) match the simulated data (solid lines) for $R$ greater than one. Note how little the value of $r_{\rm max}$ varies with cylinder radius $R$. (b) The dependence of the extent of the meniscus in the $r$ and $z$ directions on the contact angle $\theta_c$ is shown for $R=1$ and $R=10$. The simulated data (solid lines) is in good agreement with the second approximation for both values of $R$ over a wide range of contact angles, while the first approximation is in broad agreement with the data only for large $R$ and large $\theta_c$.}
\label{fig:profile3}
\end{figure}

  

\subsection{Capillary pressure}

\subsubsection{Effect of cylinder radius}

\begin{figure}
\centering
\begin{subfigure}{0.48\textwidth}
  \centering
 \includegraphics[width=\textwidth]{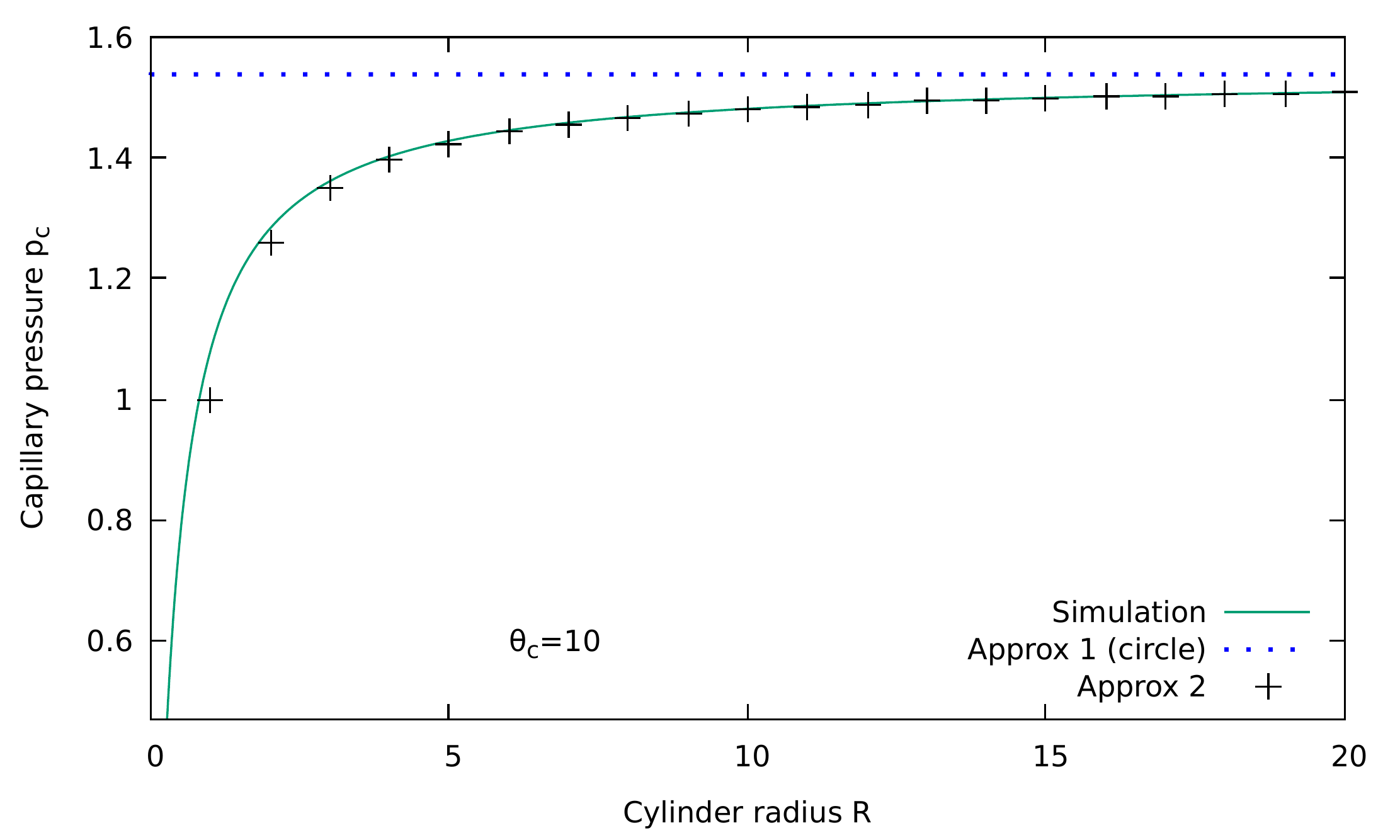}
    \caption{}
\end{subfigure}
  \begin{subfigure}{0.48\textwidth}
\includegraphics[width=\textwidth]{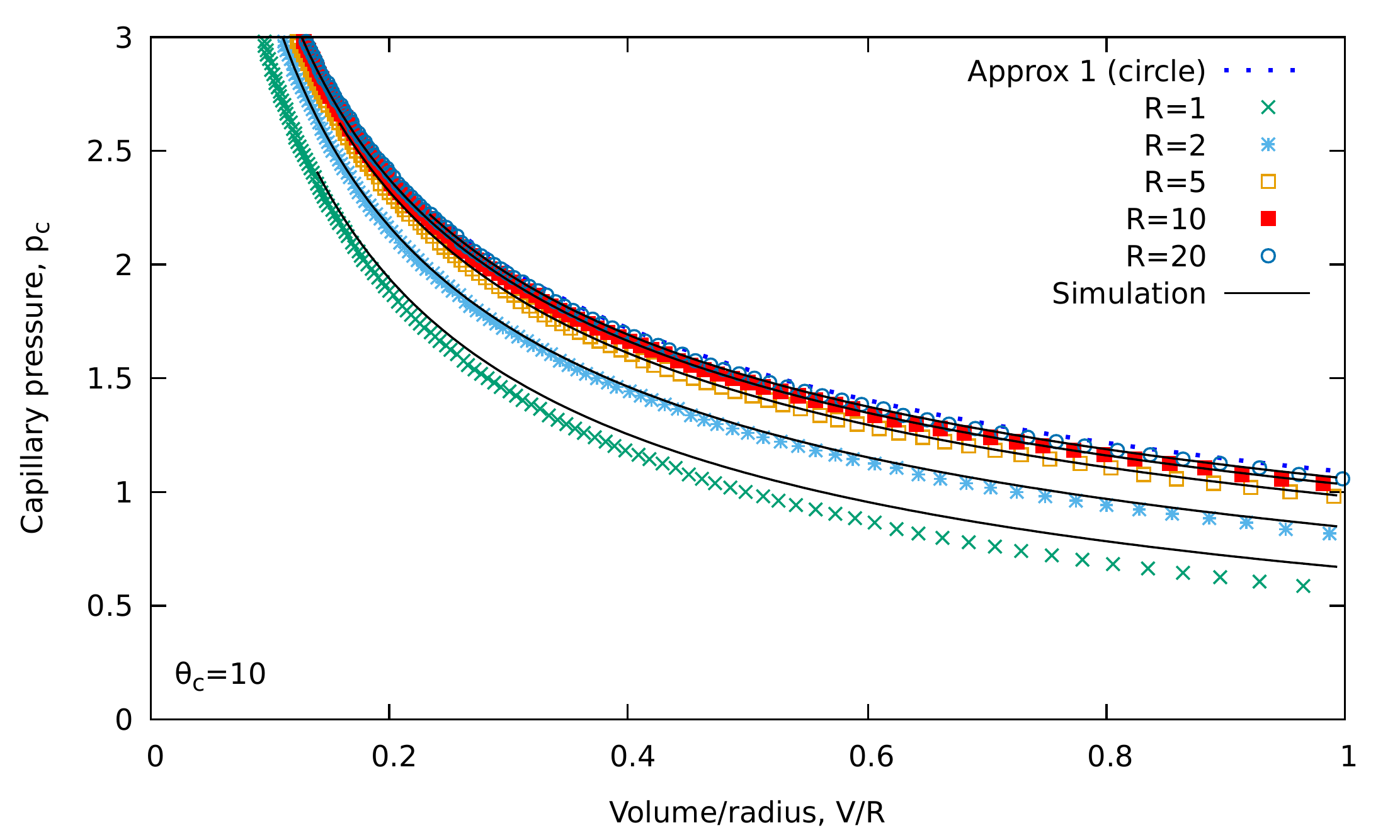}
    \caption{}
\end{subfigure}
\caption{ 
 Effect of cylinder radius on capillary pressure. (a) Comparing simulated data (solid line) with our first (horizontal dotted line) and second (points)  approximations for linearly increasing volume ($V= 0.5R$) with contact angle $\theta_c = \pi/18$.
 (b) Capillary pressure for a range of volumes $V$ and cylinder radii $R$, with contact angle $\theta_c = \pi/18$, as in Figure~\ref{fig:pc_vs_vol}(b). Simulated data is shown as solid lines, showing better agreement at larger $R$ and smaller $\theta_c$. 
}
\label{fig:p_c_volume}
\end{figure}

Figure~\ref{fig:p_c_volume}(a) shows that with liquid volume varying linearly in cylinder radius ($V = 0.5R$) and the contact angle fixed at $\theta_c = \pi/18$ the capillary pressure increases with $R$. In this respect, the capillary pressure follows a similar trend to $z_{\rm max}$:  as $R$ increases, the capillary pressure converges towards our first approximation, $\hat{p}_c \approx 1.57$. The second approximation fits the simulated data extremely well, with a small discrepancy only at small $R$.

Again for $\theta_c = \pi/18$, Figure~\ref{fig:p_c_volume}(b) shows the capillary pressure for a range of values of the liquid volume, comparing the predictions with simulated data. For each value of $R$, the capillary pressure decreases as the meniscus volume increases, most steeply for small liquid volumes and small cylinder radii. Although the first prediction works well for large $R$ (here $R=20$), the second approximation performs better down to lower values of $R$, although for large contact angles there is a deviation of a few percent in the prediction of the capillary pressure for $R=1$.

\subsubsection{Effect of contact angle}

Representative interface shapes are shown in Figure~\ref{fig:profile2}(b) for different contact angles. As the contact angle increases, the curvature reduces, and hence so does the capillary pressure. 

\begin{figure}
\centering
\begin{subfigure}{0.48\textwidth}
  \centering
 \includegraphics[width=\textwidth]{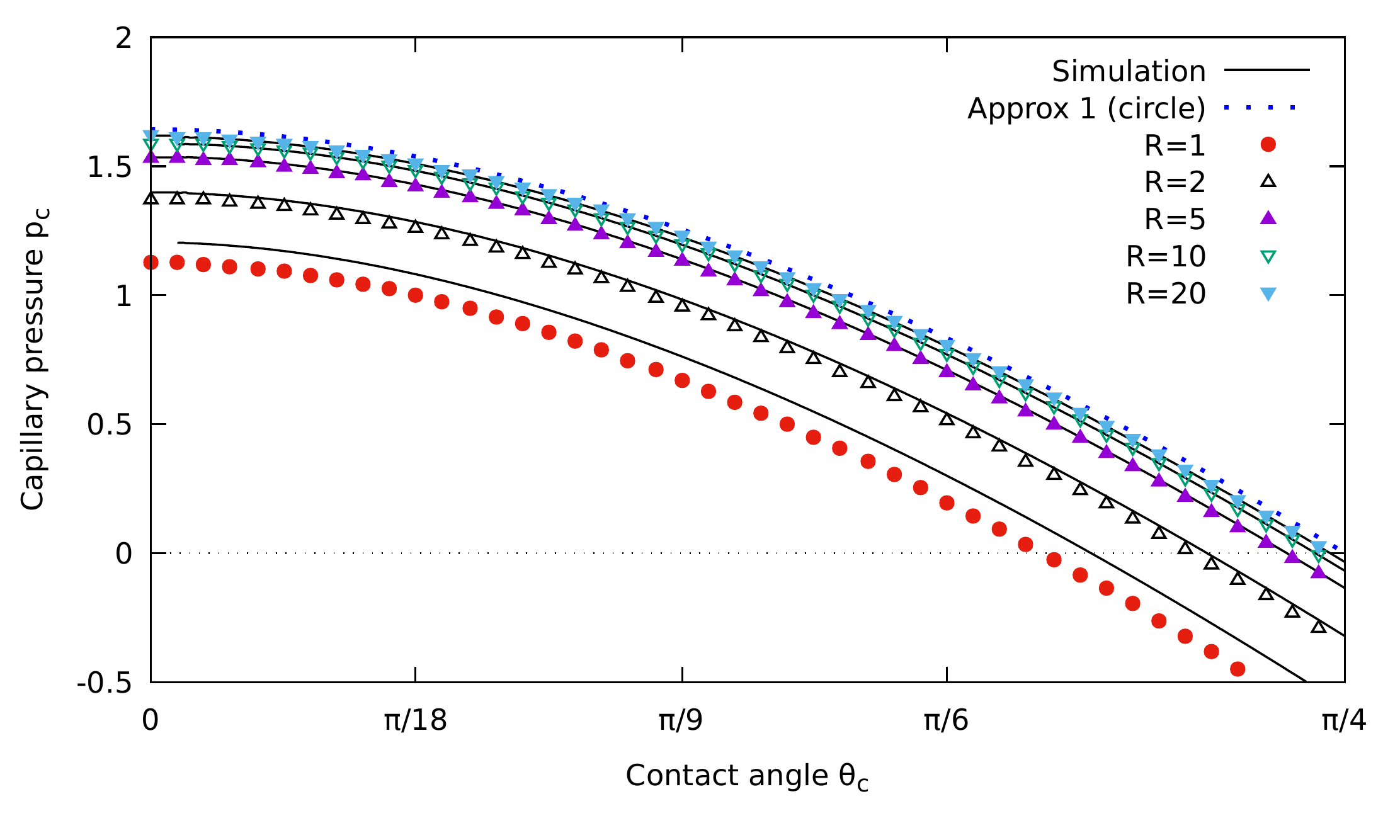}
    \caption{}
\end{subfigure}
  \begin{subfigure}{0.48\textwidth}
\includegraphics[width=\textwidth]{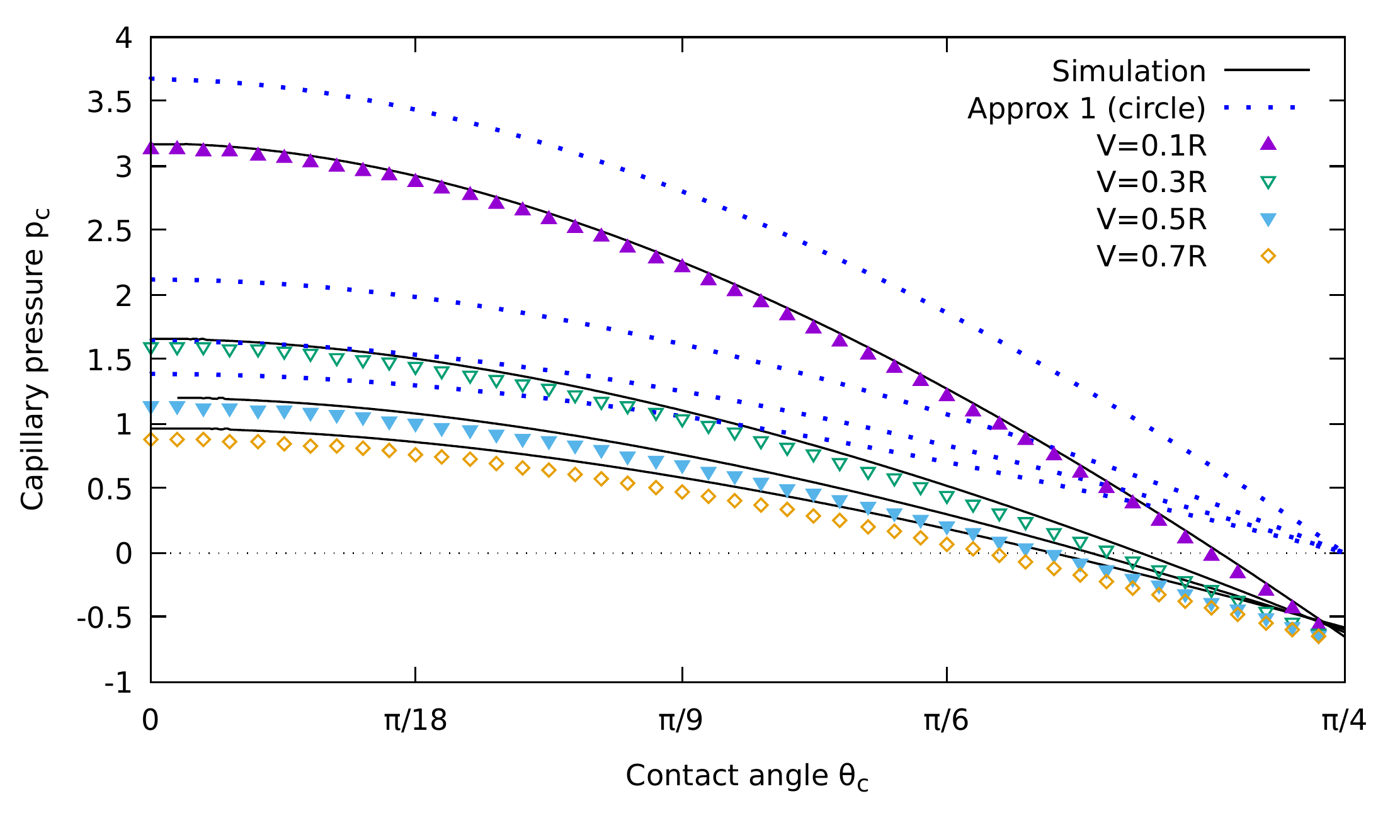}
    \caption{}
\end{subfigure}
\caption{ 
Effect of contact angle and meniscus volume on capillary pressure, comparing simulated data (solid lines) with our first (blue dotted lines) and second approximations (points).
(a) For different cylinder radii $R$ and volume $V=0.5R$.
(b) For different meniscus volumes with cylinder radius $R=1$.
}
\label{fig:p_c_angle}
\end{figure}

The same trend of decreasing capillary pressure with increasing volume is observed for all contact angles and all meniscus volumes (Figure~\ref{fig:p_c_angle})(b). The highest capillary pressures are found for the smallest contact angles, since as the contact angle increases, the interface straightens out, and for the smallest liquid volumes, since the interface is more strongly curved closer to the cylinder. For $R=1$ the second approximation works well, more so at smaller liqiud volumes / higher capillary pressures. 

Figure~\ref{fig:p_c_angle}(a) shows that our first prediction of the capillary pressure is indistinguishable from the data for all values of the contact angle for large cylinders ($R$ above about 20). The second approximation works much better at lower $R$, although for $R$ as low as one there is a discrepancy of about 10\%, almost independent of the contact angle.

For certain values of the contact angle and liquid volume, the capillary pressure (and hence the curvature of the interface) can change sign; an example is shown in Figure~\ref{fig:p_c_angle}(b) for $R=1$ and $\theta_c$ greater than $\pi/6$. The value of $\theta_c$ at which the change of sign takes place decreases as the liquid volume increases. As $R$ increases, we expect this cross-over value to converge on $\pi/4$ for any meniscus volume. Thus for sufficiently small volumes the critical contact angle at which the capillary pressure changes sign is still $\pi/4$. 

Figure~\ref{fig:p_c_angle}(b) also  suggests that there is a particular contact angle at which the capillary pressure is independent of the liquid volume for given radius $R$. This occurs at $\theta_c = \pi/4$ for large $R$, since the interface is flat (in the $(r,z)$ plane). For smaller $R$, it occurs at negative capillary pressure ($\hat{p}_c \approx =0.5$ for $R=1$), so that the interface curvature has changed sign and it bulges outwards slightly. The critical contact angle at which this convergence of the $\hat{p}_c(\theta_c)$ converges remains very close to $\theta_c=\pi/4$.

\section{Discussion and Conclusions}

We have described numerical simulations and two predictive models for the shape of the meniscus surrounding the base of a circular cylinder attached to a flat horizontal substrate. In many ways, this is simpler than the more familiar problem of the capillary rise of liquid around a cylinder immersed in a liquid~\cite{teixeira15}, and the different boundary condition at $r_{\rm max}$ allows us to make progress with  (i) a closed-form solution that works well at large $R$, i.e. cylinders that are many times wider than their height, and (ii) a semi-analytic solutions that works well for a much broader range of values of $R$, showing significant error only at large liquid volumes (when gravity might be anticipated to have a further significant effect) and small $R$ less than one. In this second approximation, Figure~\ref{fig:pc_vs_vol} shows how the capillary pressure depends on liquid volume, removing the need for numerical calculations, with eqns.~(\ref{eq:rtheta}) and (\ref{eq:ztheta}) giving the shape of the meniscus.

Notable features of the solutions we give are:
\begin{itemize}
    \item The radial extent of the meniscus is much better described than the meniscus height by our first approximation, while the second approximation captures both well; indeed, the radial extent of the meniscus appears almost independent of cylinder radius (Figure~\ref{fig:profile3}(a)) down to $R$ below $1$ (in units of the channel depth);
    \item the critical contact angle at which the interface curvature/capillary pressure changes sign decreases as the cylinder radius decreases and as the meniscus volume increases (Figure~\ref{fig:p_c_angle});
    \item there appears to be a particular contact angle at which the capillary pressure is the same, irrespective of the meniscus volume, for given cylinder radius (Figure~\ref{fig:p_c_angle}(b)).
\end{itemize}

By establishing predictive models for the capillary pressure in this idealized geometry, we aim to  provide quantitative results to enable comparison with experiment. Providing results for capillary pressure and interface shape for a broader range of geometries, for example cylindrical pillars with small radius $R$, different shapes of pillars, or more than one pillar, is the subject of ongoing research.

\section*{Acknowledgments}

SC thanks W.R. Rossen, M.I.C. Teixeira and P.I.C. Teixeira for useful suggestions and K. Brakke for the development of the Surface Evolver. An AberDOC PhD scholarship from Aberystwyth University is gratefully acknowledged.

\end{document}